\documentclass[twocolumn,prl,showpacs,superscriptaddress]{revtex4}
\usepackage{graphicx}
\usepackage{amsmath}
\usepackage{amssymb}

\begin{document}
\title{Quantum Spin Hall Effect in Two-dimensional Crystals \\
of Transition Metal Dichalcogenides}
\author{M.~A. Cazalilla}
\affiliation{Department of Physics, National Tsing Hua University, 
and National Center for Theoretical Sciences (NCTS), Hsinchu City, Taiwan.}
\affiliation{Donostia International Physics Center (DIPC),  Manuel de Lardizabal 4, 
20018, San Sebastian, Spain.}
\author{H. Ochoa}
\author{F. Guinea}
\affiliation{Instituto de Ciencia de Materiales de Madrid (ICMM). CSIC, Cantoblanco. E-28049 Madrid. Spain.}
\date{\today}
\begin{abstract}
We propose to engineer time-reversal-invariant topological insulators in two-dimensional (2D) crystals of transition metal dichalcogenides (TMDCs).  We note that, at low doping, semiconducting TMDCs under shear strain will develop spin-polarized Landau levels residing in different valleys.   We argue that gaps between Landau levels in the range of $10-100$ Kelvin are within experimental reach. In addition, we point out that a superlattice arising from a Moir\'e pattern  can lead to topologically non-trivial subbands. As a result, the edge transport becomes quantized, 
which can be probed in multi-terminal devices made using strained 2D crystals and/or heterostructures. The strong $d$ character of valence and conduction bands may also allow for the investigation of the effects of electron correlations on the topological phases.
\end{abstract}
\maketitle
%

%\section{Introduction}
%
There is currently much interest in
two-dimensional (2D) crystals of transition metal dichalcogenides
(TMDC) such as MoS$_2$ or WSe$_2$~\cite{dical}.
Compared to Graphene, many of
these materials are semiconductors with
sizeable gaps ($\approx 1$ eV),
which makes them good candidates
for applications  in conventional electronics  such
as the manufacture of transistors~\cite{transistor}.

 By contrast,  the quantum spin Hall (QSH)
effect~\cite{cite_qshe}  has been postulated as the paradigm
for future \emph{non-dissipative}
nano-electronics~\cite{qshe_app}.
Indeed,  provided that magnetic impurities
(or other time-reversal symmetry-breaking perturbations)
can be eliminated, electron transport through the
edge of a QSH insulator becomes ballistic.  This is because
counter propagating channels of the same spin are
spatially separated.

 Nevertheless,  the QSH effect has been
observed in very few materials~\cite{mercury_tell,other} so far,
making the search for new  systems exhibiting this
effect a major scientific endeavor. In the context
of 2D crystals, Weeks and coworkers have recently
proposed an approach to enhance the spin-orbit
coupling in graphene by heavy ad-atom deposition~\cite{Weeks},
which could lead to a realization of the Kane-Mele
model~\cite{kane_mele} and the QSH effect. The availability of
2D  materials exhibiting this effect may 
allow for the construction of  flexible electronic devices 
with low-power consumption.  

In this work, we present a proposal for engineering the QSHE in
strained 2D crystals and heterostructures made using TMDCs.
Strain is known to induce pseudo-magnetic fields in 2D
crystals~\cite{vozmediano_katsnelson_guinea,guinea_katsnelson_geim, crommie}. These
fields have been predicted~\cite{guinea_katsnelson_geim} and experimentally found~\cite{crommie}
to produce Landau levels (LLs) in the
electronic spectrum of 2D crystals such as graphene. As explained below, it is also possible to
exploit these
pseudo-magnetic fields to engineer time-reversal invariant
topological phases in other 2D crystals such as the TMDCs.
This approach has several attractive features. Using
MoS$_2$ as an example, we find that gaps
between LLs scale as $\hbar\omega_c/k_B \simeq 2.7~B_0[T]~$K, where
$B_0[T]$ is the strength of the pseudo-magnetic field in Tesla.
Fields $B_0[T] \sim 10-10^2$ T have been experimentally 
demonstrated~\cite{guinea_katsnelson_geim,crommie},
which in the case of TMDCs could lead to LL gaps of up to $\approx 100$ K. By comparison,
the gaps achievable by straining GaAs
are in the range of tens of mili-Kelvin~\cite{bernevig_zhang}. 
The much larger magnitude of the LLs gaps in the present proposal stems 
the strain coupling to the electron current, rather than the electron spin current as in Ref.~\cite{bernevig_zhang}. These two coupling constants are, in general, vastly different in order of magnitude. However, in
multi-valley systems like graphene, strain only leads to (spin) unpolarized LLs. What sets semiconducting TMDCs apart is the spin-orbit coupling (SOC) that produces a large spin splitting of the valence (and to a smaller extent, the conduction) band.  For small doping, this leads to spin-polarized LLs in different valleys, which opens the 
possibility of realizing time-reversal invariant (TRI) topological phases.
Furthermore,  the valence and conduction band of semiconducting TMDC have strong 
$d$ character~\cite{two_band}, which, along with  the poor screening of the 
Coulomb interaction in 2D, means that  
electron interactions  can have interesting effects on properties 
of the TRI topological phases realized in TMDCs.
%
%\section{Model}

\begin{figure}[t]
\includegraphics[width=\columnwidth]{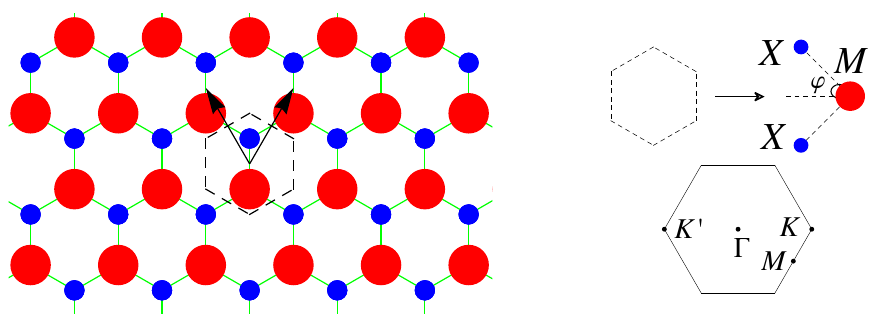}\\
 \caption{(Color Online) Top view of the 2D lattice of TMDC. The unit cell and Brillouin zone are shown in the right panel. $M$ corresponds to the transition metal atom, whereas $X$ to the chalcogen atoms.}
\label{fig:fig2}
\end{figure}

Bulk TMDC are composed of $X$-$M$-$X$ layers stacked on top of each other and coupled by weak Van der Waals forces. Therefore, like graphite, these materials can be exfoliated down to a single  layer. The transition metal atoms ($M$) are arranged in a triangular lattice and each one is bonded to six chalcogen atoms ($X$) (see Fig.~\ref{fig:fig2} for a top view of the lattice). The point group of the 2D crystal is $D_{3h}$. The Fermi level lies at the two inequivalent corners of the Brillouin zone, $K$ and $K^{\prime}$ points, as in the case of graphene, but the spectrum is gapped. The character of the conduction and valence bands is dominated by $d_{3z^2-r^2}$ and $d_{x^2-y^2}\pm id_{xy}$ orbitals from $M$ atom, respectively, where the sign $+$ ($-$) holds for $K$ ($K^{\prime}$) point. Furthermore,  both bands have a non-negligible hybridization with the bonding combination of $p_x\pm ip_y$ (in the conduction band) and $p_x\mp ip_y$ (in the valence band) orbitals from X atoms.

In the continuum limit, the system is well described by the  following two-band Hamiltonian~\cite{two_band,tb,ochoa_roldan}:
\begin{equation}
H_0 = v (\tau^z \sigma^x p_x + \sigma^y p_y) + \Delta \sigma^z +
\frac{\lambda_{\mathrm{SO}}}{2} \tau^z s^z\left(1 - \sigma^z \right), \label{eq:ham}
\end{equation}
where $\mathbf{p} = (p_x, p_y)$ is the electron crystal
momentum referred to the $K$ ($K^{\prime}$) point,
for which we introduce $\tau^{z} = +1$  ($\tau^{z} = -1$); $\sigma^{\alpha}$ are
Pauli matrices acting on the space span by conduction and valence band states; $\Delta$ is the band gap and  $v = t a/\hbar$,
 where $a$ is the lattice parameter and $t$ is a phenomenological hybridization parameter;   finally, $s^z$ is the spin projection along the
$z$ axis perpendicular to the 2D crystal.
The Hamiltonian \eqref{eq:ham} can be constructed phenomenologically by considering scalar invariants of $D_{3h}$ formed from operators $\sigma^{\alpha}$, $\mathbf{p}$ and spin $s^{\alpha}$. The symmetry properties of these operators are listed in Table~\ref{table1}. Besides the gap, another important difference
with graphene is the spin splitting of the bands due to the strong spin-orbit interaction provided by $M$ atoms. Below, we focus on p-doped TMDCs, which allows us to neglect the smaller spin splitting of the conduction band~\cite{ochoa_roldan}. For the valence band, the latter is in the range of $100$-$400$ meV depending on the material~\cite{so_dft,arpes,spinorbit}. The Hamiltonian~\eqref{eq:ham} can be also derived from a microscopic theory.
From a simplified tight-binding model~\cite{epaps}, we obtain $\Delta=1.41$ eV and $t=1.19$ eV.
\begin{table}[t]
\begin{tabular}{|c|c|c|}
\hline
Irrep & TR Even & TR Odd \\
\hline
\hline
$A^{\prime}_1$ & $1, \sigma^z$, $\sum_{\alpha} u_{\alpha\alpha}$ & - \\
\hline
$A^{\prime}_2$ & - & $\tau^z,s^z$ \\
\hline
$E^{\prime}$ & $(\sigma^x, \tau^z \sigma^y)$, $(u_{xx}-u_{yy},-2u_{xy})$ &  $(\tau^z \sigma^x, \sigma^y)$, $\mathbf{p}=(p_x,p_y)$\\
\hline
$E^{\prime\prime}$ & - & $(s^x, s^y)$ \\
\hline
\end{tabular}
\caption{\label{table1}Symmetry classification of the electronic operators and strain tensor components according to the irreducible representations (Irrep) of $D_{3h}$ and time reversal operation.}
\end{table}

In order to turn a TMDC into a TRI
topological insulator, we need to consider the effect of strain
on the band structure. Strain
is described by a rank-$2$ tensor,
$u_{\alpha\beta} =  \frac{1}{2} (\partial_{\alpha} u_{\beta} + \partial_{\beta}
u_{\alpha} + \frac{\partial h}{\partial x_{\alpha}} \frac{\partial h}{\partial x_{\beta}})$, where $\mathbf{u} = (u_x, u_y)$ ($h$) is the  in-plane (out-of-plane)  displacement of the unit cell in the  long-wave length limit. The three components of
$u_{\alpha\beta}$ can be split according to irreducible
representations of $D_{3h}$ as shown in Table~\ref{table1}.
Thus, the coupling with strain reads:
\begin{align}
H_{\mathrm{strain}} = \beta_0 t \sum_{\alpha}u_{\alpha\alpha} + \beta_1 t \sum_{\alpha}u_{\alpha\alpha}
\sigma^z
\nonumber\\
+ \beta_2 t \left[\left(u_{xx}-u_{yy}\right)\sigma^x-2u_{xy}\tau^z\sigma^y\right]. \label{eq:ham_strain}
\end{align}
Microscopically, the origin of these couplings is the change in the hybridization between $d$ orbitals from $M$ atoms and $p$ orbitals from $X$ atoms due to the distortion of the lattice. In our microscopic tight-binding model~\cite{epaps}, the phenomenological constants $\beta_{0,1,2}$ are given by the Gr\"uneisen parameters~\cite{vozmediano_katsnelson_guinea} associated to the hopping amplitudes considered in the calculation~\cite{epaps}.
The trace of the strain tensor generates scalar potentials of different
strength in the valence and conduction bands. In addition, strain can be introduced
in  Eq.~\eqref{eq:ham} as a minimal coupling  $\mathbf{p} \to \mathbf{p} - e\mathbf{A}$  to a vector potential
$e\mathbf{A} = \frac{\hbar\beta_2}{a} \tau^z
(u_{yy}-u_{xx}, 2 u_{xy})$.
The presence of
$\tau^z$ indicates that the pseudo-magnetic field
has opposite sign on different valleys, which is necessary as
strain does not violate TRI. 

Henceforth, we assume that the system is doped with holes and therefore the Fermi level crosses the valence band. Integrating out the conduction band to leading order in $\Delta^{-1}$ yields:
\begin{equation}
H_{v} = -\frac{\Pi_{+}(\tau^z)\Pi_{-}(\tau^z)}{2m^*} + U(\mathbf{r}) + \lambda_{\mathrm{SO}} s^z \tau^z,\label{eq:ham2}
\end{equation}
where $m^* = \Delta/2v^2 \simeq 0.5$~\cite{eff_mass}, $U(\mathbf{r}) = g (u_{xx}+ u_{yy})$ with $g=t\left(\beta_0-\beta_1\right)$, and $\Pi_{\pm}(\tau^z) = (\tau^z p^x \pm i p^y) - e (\tau^z A_x \pm i A_{y})$,
which obey the commutation relation $\left[ \Pi_{+}(\tau^z), \Pi_{-}(\tau^z)
\right]  =  2 e\hbar \tau^z  B(\mathbf{r})$, where $B(\mathbf{r}) = \partial_x A_y  - \partial_y A_x$. Corrections of $\mathcal{O}\left(\Delta^{-2}\right)$
have been also obtained and can lead to mixing of the Landau levels, but they can
be  neglected as for typical parameters
$\hbar \omega_c  \lesssim 10^{-2} \Delta$.

\begin{figure}[t]
\includegraphics[width=\columnwidth]{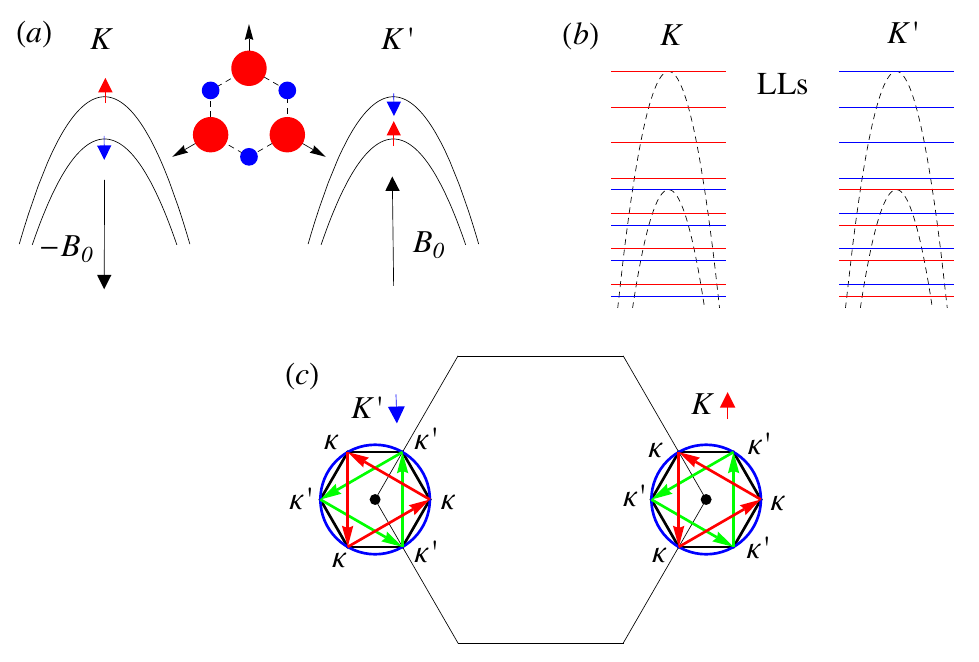}\\
 \caption{(Color Online) (a)
 Low-energy spectrum of a semiconducting transition metal dicalchogenide around the $K$ and $K^{\prime}$ points of the first Brioullin zone described by the continuum-limit Hamiltonian in Eq.~\eqref{eq:ham2}. The inset shows the orientation of the stress tensor field discussed through the text with respect to the lattice. (b) Schematic representation of the Landau Levels (LLs) induced in the valence band when strain is applied. (c) A superlattice can be also used to create a topologically non-trivial
subband structure. A triangular superlattice produces replicas of $K$ and $K^{\prime}$ by mixing the valence and conduction band states at the mini-Dirac points as indicated by the arrows.}  
 \label{fig:fig1}
\end{figure}

 In the absence of strain (i.e. $\mathbf{A} = U =  0)$,
 Eq.~\eqref{eq:ham2} describes the Bloch states at the top of the valence band near
 $K,K^{\prime}$. Owing to the spin-orbit coupling ($\propto \lambda_{\mathrm{SO}}$),
for small hole doping (i.e. $|\epsilon_F| \ll \lambda_{\mathrm{SO}}$), the spin and valley spin of the holes are locked to each other, i.e. only holes with either $(K,\uparrow)$ or $(K^{\prime},\downarrow)$  can exist (cf. Fig.~\ref{fig:fig1}). Furthermore,
the states at  $K$ and $K^{\prime}$ are Kramers pairs. This feature is crucial for the realization of a time-reversal
invariant topological phase, as we argue below.

%\section{Landau Levels}

 Applying a pure shear strain, i.e.
$u_{xx} + u_{yy} = 0$ (i.e. $ U = 0$), and
$u_{xx} = - u_{yy} = - C y$,
$u_{xy}  = - Cx$, we have $B_z(\mathbf{r})  = \partial_x A_y
- \partial_y A_x = - \tau^z B_0$ with $B_0=\frac{4C\hbar\beta_2}{ea} > 0$~\cite{beta}. The Hamiltonian
in \eqref{eq:ham2} can be diagonalized after introducing
$a_{K} = \Pi_{-}(\tau^z = +1)/\sqrt{2 e \hbar B_0}$
and $a_{K^{\prime}} = \Pi_{-}(\tau^z = -1)/\sqrt{2 eB_0}$:
\begin{equation}
H = -\hbar \omega_c \left( a^{\dag}_K a_{K} + a^{\dag}_{K^{\prime}} a_{K^{\prime}}\right) +  \lambda_{\mathrm{SO}} s^z \tau^z,
\end{equation}
where $\omega_c  = e B_0/m^*$. As stated
above, even for the largest achievable pseudo-magnetic fields
($B_0\sim 10^2~$T) $\hbar \omega_c \ll \lambda_{\mathrm{SO}}$,
and therefore,  provided $
|\epsilon_F| \ll \lambda_{\mathrm{SO}}$,
LLs at different valleys
are occupied by holes with opposite
spin. The resulting model has been shown by
Bernevig and Zhang~\cite{bernevig_zhang} to display
the QSH effect, meaning that  the Hall conductivity
 $\sigma^{xy}_H$ is zero but the spin-Hall conductivity $\sigma^{xy}_{sH}$ is quantized in units of $2 e/4\pi$.

 The strain configuration described above
can be  created by the methods  described in Ref.~\cite{guinea_katsnelson_geim}. For a TMDC 2D crystal
flake under trigonal strain~\cite{guinea_katsnelson_geim}
(cf. Fig.~\ref{fig:fig1}),
one concern is the sample  size. The latter is limited by the
maximum tensile strength of the TMDCs,
$T_{\mathrm{max}}$. For MoS$_2$,
the thermodynamic relation $\sigma_{\alpha\beta}=
2\mu u_{\alpha\beta}$ ($\sigma_{\alpha\beta}$ being
the stress tensor and $\mu$ is the shear modulus) allows us
to estimate the maximum sample size
$L\approx \frac{2 T_{\mathrm{max}}}{\mu C}$. The values of
$\mu \simeq 50.4$ N/m and
$T_{\mathrm{max}} \simeq 16.5$ N/m can be obtained from
density-functional calculations~\cite{dft}.
This yields a relation between the maximum pseudo-magnetic
field (in Tesla) and the sample size $L$ in $\mu\mathrm{m}$: $B_0[T] \approx 8/L[\mathrm{\mu m}]$. Using $\hbar\omega_c/k_B = 2.7 \, B_0[T]$ and taking
$L \approx 1\: \mathrm{\mu m}$, we estimate
$\hbar\omega_c/k_B \simeq 20$ K for MoS$_2$.

 For small strained 2D crystal flakes,
 it is necessary to  take into account the effect of an inhomogeneous pseudo-magnetic field resulting from a non-uniform
strain distribution. In this regard, we
note that the lowest
LL eigenfunctions are null eigenvectors
of $\Pi_{-}(\tau^z)$:
\begin{equation}
\Pi_{-}(\tau^z)\psi(\mathbf{r}) = 0, \label{eq:lll}
\end{equation}
Therefore, following~\cite{Aharonov}, we write
$\mathbf{A}(\mathbf{r}) = \tau_z \left(\mathbf{\hat{z}} \times \boldsymbol{\nabla}\chi(\mathbf{r}) + \boldsymbol{\nabla}\phi(\mathbf{r})\right)$, which allows
to solve \eqref{eq:lll} for $K$ [$K^{\prime}$] and yields $\psi(\mathbf{r})
= f(z^*) e^{\frac{2\pi}{\Phi_0} \left(\chi(\mathbf{r})+i\phi(\mathbf{r})\right)}$ [$\psi(\mathbf{r})
= f(z) e^{\frac{2\pi}{\Phi_0} \left(\chi(\mathbf{r})-i\phi(\mathbf{r})\right)}$], where
$f(z^*)$ [$f(z)$] is a polynomial of
$z = (x + i y)$ [$z^*= (x - i y)$]
of maximum degree $N = [\Phi/\Phi_0]$,
$\Phi = \int d\mathbf{r}\:  B_0(\mathbf{r})>0$ being the total flux
and $\Phi_0 = h /e$ the flux quantum~\cite{note}. Hence, the wave-function
describing  $N_{\uparrow} = N_{\downarrow} = N$ (non-interacting) electrons in the lowest
LL reads~\cite{Aharonov}:
\begin{equation}
\Phi_0(\{\mathbf{r}_{i\alpha}\}) = e^{-F}
\prod_{i<j}  (z^*_{i \uparrow} - z^*_{j \uparrow}) (z_{i\downarrow} - z_{j\downarrow}),
\end{equation}
where $F(\{\mathbf{r}_{i\alpha}\}) = \frac{2\pi}{\Phi_0}
\sum_{i=1,\alpha=\uparrow,\downarrow}^N \left[ \chi(\mathbf{r}_{i\alpha})  -i\alpha\phi(\mathbf{r}_{i\alpha})\right]$.

% \section{MOIRE ALTERNATIVE}

 Larger sample sizes can be achieved by other methods such in 2D crystal bubbles ~\cite{guinea_katsnelson_geim,crommie}. A periodic array of such bubbles  will lead to periodic modulation of strain and pseudo-magnetic field, which  allows  to  create topologically non-trivial band structures~\cite{guinea_katsnelson_geim,lacroix}. Alternatively,  a superlattice can be used to create, within each valley, a band with non trivial topological properties. If we neglect trigonal warping, each valley has a time-reversal-like symmetry where $\Psi_{A/B} ( \vec{\bf r} ) \rightarrow \Psi_{A/B}^* ( \vec{\bf r} )$. This symmetry is broken by "gauge" terms proportional to $\sigma_x$ and $\sigma_y$ in Eq.~\eqref{eq:ham}, which can arise from strain or from virtual hopping processes to the substrate~\cite{epaps}. A suitable combination of periodic scalar and a gauge potentials can~\cite{LGK11}: i) separate the lowest leading subbands from the rest by opening a gap at the edges of the superlattice Brillouin Zone (see Fig.~\ref{fig:fig1}), and ii) give a total Chern number of $\pm 1$ to the subbands arising at the $K , K'$ valleys, respectively. This approach, which relies on a periodic magnetic field with zero average leads to a Quantum Hall insulator~\cite{S09} in the absence of a global magnetic flux~\cite{H88}. The periodicity of the potential should be such that the width of the subbands is smaller than the spin splitting in each valley. Large enough periodicities where subbands can be resolved have been achieved for graphene on 
Boron Nitride~\cite{Yetal12,Petal13,Detal13,Hetal13}. A different scheme leading to the QSH effect based on a Moir\'e pattern in a single valley semiconductor like GaAs is suggested in~\cite{SN13}.

%\section{EXPERIMENTAL CONSEQUENCES}

We next discuss some experimental
consequences of our predictions.
As mentioned above, the Hall conductivity
 $\sigma^{xy}_H$ is zero but the spin-Hall conductivity $\sigma^{xy}_{sH}$ is quantized in units of $2 e/4\pi$~\cite{bernevig_zhang}. However, if the total $s^z$ is not a good  quantum number (see discussion below), $\sigma^{xy}_{sH}$ is not
 exactly quantized~\cite{cite_qshe}. Instead,  charge transport through
 the (helical) edge channels  provides a clearer signature of existence
 of a topological phase~\cite{mercury_tell,other}.

 Nevertheless, in the case of strained crystals in the
 absence of a superlattice potential,  we must be careful
in qualifying the  strained 2D crystal as a topological insulator
for arbitrary LL filling. This is because adatoms, a perpendicular electric field, out of plane deformations, etc. break the mirror symmetry
about the 2D crystal plane, which induces a Rashba spin-orbit
coupling $~\sim (\tau^z\sigma^x s^y -
\sigma^y s^x)$ in the Hamiltonian of Eq.~\eqref{eq:ham}.
Rashba allows for spin flips and therefore can lead
to backscattering between counter-propagating edge channels.
For an odd number  of occupied LLs, an
odd number of Kramers' pairs of
edge  modes cross the Fermi energy and,
for weak to moderate electron-electron interactions, the integrity of
at least one Kramers' pair of edge modes against TRI perturbations that induce spin-flip scattering like
Rashba spin-orbit coupling is always ensured~\cite{xu_moore}.
Thus, for $2n+1$ (with $n$ integer)
occupied LLs, the system is TRI protected topological phase and
a two-terminal measurement of the conductance will
yield at least $2 e^2/h$ and at most $2 (2n+1) e^2/h$, depending
of degree of edge disorder and other $s^z$ non-conserving perturbations.

 On the other hand, if the number of occupied LLs is even ($=2n$), there
will an even number of pairs of edge modes crossing the Fermi level and this situation is no longer protected against e.g. Rashba-type disorder
potential~\cite{xu_moore} (although edge modes  survive for strong enough electron-electron interactions~\cite{xu_moore}). However, in sufficiently clean samples and provided interactions are weak, quantized  conductance of $ 4n e^2/h$  may be observable. Furthermore, the existence of bulk LLs can be detected by means
of scanning tunneling microscopy as in the case of graphene~\cite{crommie}.

Finally, let us discuss the possible effect of interactions.
The strong $d$ character of the valence and conduction bands
means that electron correlations can have a important effect
on the topological phases, especially on the
edge states~\cite{xu_moore,iridates}. Indeed, for MoS$_2$
the short-range part of the interaction (i.e. the Hubbard-$U$)
has been estimated in Ref.~\cite{roldan} to be $\sim 2$-$10$ eV.
Thus,  MoS$_2$ may present a scenario comparable
to the Iridates~\cite{iridates}. However, the QSH effect in
the TMDCs may allow for a more complete understanding of the interplay
between electron correlation and QSH physics, since correlation effects decrease
as the metal atom $M$ is varied from the $4d$ series (as in MoS$_2$) to $5d$
series (as in WS$_2$).

%\section{OUTLOOK AND CONCLUSIONS}

  In conclusion, we have presented a proposal to engineer
time-reversal invariant topological phases in 2D crystals of transition
metal dicalchogenides (TMDC) under strain and/or in heterostructures which
induce superlattice potentials. We also note that the proposal
can be extended to electron-doped TMDC. However,
in this case, the separation between the Landau levels ($\hbar \omega_c/k_B \simeq 2.2 \: B_0[T]$)
and the strength of the superlattice potential are limited by the much smaller spin-orbit coupling
($\lambda_{\mathrm{SO}}\sim 10 \: \mathrm{meV} \sim 100$ K)~\cite{so_dft}, which  also
requires much lower temperatures. We
hope that the feasibility of our proposal
will stimulate further experimental work along these
directions. We stress that compared to other QSH
systems~\cite{mercury_tell,other}, strained 2D TMDCs
allow for much larger tunability of material parameters
as well as the strength of the Berry curvature responsible for
the Landau levels.  In this regard, the main obstacle for the
observation of the QSH effect appears to be the reduced
carrier mobility in currently available 2D crystals of TMDCs.
However, given the notorious technological potential of
these materials, we  expect this obstacle will be
overcome in the near future.

MAC acknowledges financial support from a start-up fund
from NTHU and NCS and NCTS (Taiwan). HO acknowledges 
financial support through a JAE-Pre grant (CSIC, Spain). FG acknowledges
support from  the Spanish Ministry of Economy (MINECO)
through Grant no. FIS2011-23713, the European Research Council
Advanced Grant (contract 290846) and from European
Commission under the Graphene Flagship contract
CNECT-ICT-604391.

\clearpage
\begin{widetext}

\section{Supplementary information}

\subsection{Microscopic derivation of the two-bands Hamiltonian}

\subsubsection{Tight-binding model}

We consider a simplified tight-binding Hamiltonian acting on the subspace span by the symmetry-adapted Bloch wave functions $\left|\mbox{M }d_{3z^2-r^2}\right\rangle$, $\left|\mbox{M }d_{x^2-y^2}+\tau id_{xy}\right\rangle$, $\left|\mbox{X (b) }p_{y}+i\tau p_x\right\rangle$, and $\left|\mbox{X (b) }p_{y}-i\tau p_x\right\rangle$. The model is strictly valid around $\mathbf{K}_{\tau}$ points, where $\tau=\pm 1$ labels the two valleys $K$ ($\tau = +1$) and $K^{\prime}$ ($\tau = -1$). The crystal field parameters are:
\begin{gather}
\left\langle \mbox{M }d_{3z^2-r^2} \right|\mathcal{H}_{TB}\left| \mbox{M }d_{3z^2-r^2}\right\rangle=\Delta_0\nonumber\\
\left\langle \mbox{M }d_{x^2-y^2}+\tau id_{xy} \right|\mathcal{H}_{TB}\left| \mbox{M }d_{x^2-y^2}+\tau id_{xy}\right\rangle=\Delta_2\nonumber\\
\left\langle \mbox{X (b) }p_y\pm\tau ip_{x} \right|\mathcal{H}_{TB}\left| \mbox{X (b) }p_{y}\pm\tau ip_{x}\right\rangle=\Delta_p.
\end{gather}
The hopping integrals sketched in Fig.~\ref{fig:hoppings} can be expressed in terms of the two-center Slater-Koster parameters $V_{dp\sigma}$, $V_{dp\pi}$~\cite{Slater-Koster} as follows:
\begin{gather}
t_x=\frac{\sqrt{3}}{2}V_{pd\sigma}\cos\varphi\nonumber\\
t_y=-V_{pd\pi}\cos\varphi \nonumber\\
t_z=\cos\varphi\left(\sin^2\varphi-\frac{1}{2}\cos^2\varphi\right)V_{pd\sigma}-\sqrt{3}\cos\varphi\sin^2\varphi V_{pd\pi},
\end{gather}
where the angle $\varphi$ is defined in the main text. The matrix elements between Bloch states of orbitals of M and X atoms at $\vec{q}$ read:
\begin{gather}
\left\langle \mbox{M }d_{3z^2-r^2} \right|\mathcal{H}_{TB}\left| \mbox{X (b) } p_y\pm i\tau p_x\right\rangle=\sum_{\hat{\delta}}e^{-ia\vec{q}\cdot\hat{\delta}}t_{z}\left(\hat{y}\pm i\tau\hat{x}\right)\cdot\hat{\delta}
\nonumber\\
\left\langle \mbox{M }d_{x^2-y^2}+\tau id_{xy}\right|\mathcal{H}_{TB}\left| \mbox{X (b) } p_y\pm i\tau p_x\right\rangle=
\sum_{\hat{\delta}}e^{-ia\vec{q}\cdot\hat{\delta}}
t_{x}\left(\hat{y}\pm i\tau\hat{x}\right)\cdot\hat{\delta}\times\left[2\left(\hat{x}\cdot\hat{\delta}\right)^2-2i\tau\left(\hat{xy}\cdot\hat{\delta}\right)^2+i\tau-1\right]+
\nonumber\\
+\sum_{\hat{\delta}}e^{-ia\vec{q}\cdot\hat{\delta}}t_{y}\left(\hat{y}\pm i\tau\hat{x}\right)\cdot\hat{\delta}_{\perp}\times\left[2\left(\hat{\delta}\cdot\hat{x}\right)
\left(\hat{\delta}_{\perp}\cdot\hat{x}\right)-2i\tau\left(\hat{\delta}\cdot\hat{xy}\right)
\left(\hat{\delta}_{\perp}\cdot\hat{xy}\right)\right]
\end{gather}
where the unit vectors are defined as:\begin{gather}
\hat{x}=\left(1,0\right)\nonumber\\
\hat{y}=\left(0,1\right)\nonumber\\
\hat{xy}=\frac{1}{\sqrt{2}}\left(1,1\right)\nonumber\\
\mbox{and }\hat{\delta}_{\perp}=\left(\delta_y,-\delta_x\right)
\end{gather}
and $\hat{\delta}=\left(\delta_x,\delta_y\right)$ are the three unit vectors connecting nearest-neighbors in the honeycomb lattice:
\begin{equation}
\hat{\delta}=\left\{\hat{y},\left(\frac{\sqrt{3}}{2},-\frac{1}{2}\right),\left(-\frac{\sqrt{3}}{2},-\frac{1}{2}\right)\right\}
\end{equation}
\begin{figure}
\includegraphics[width=0.35\textwidth]{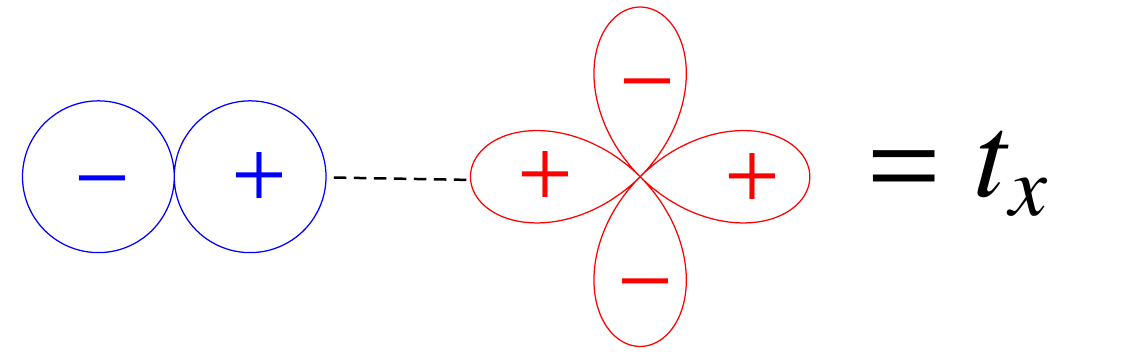}
\includegraphics[width=0.35\textwidth]{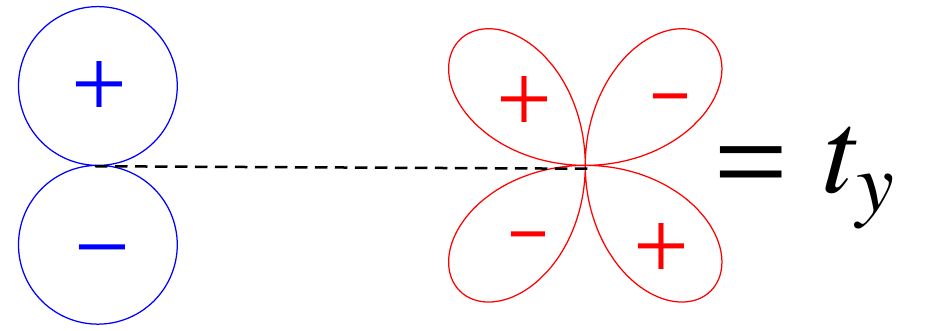} \\
\includegraphics[width=0.3\textwidth]{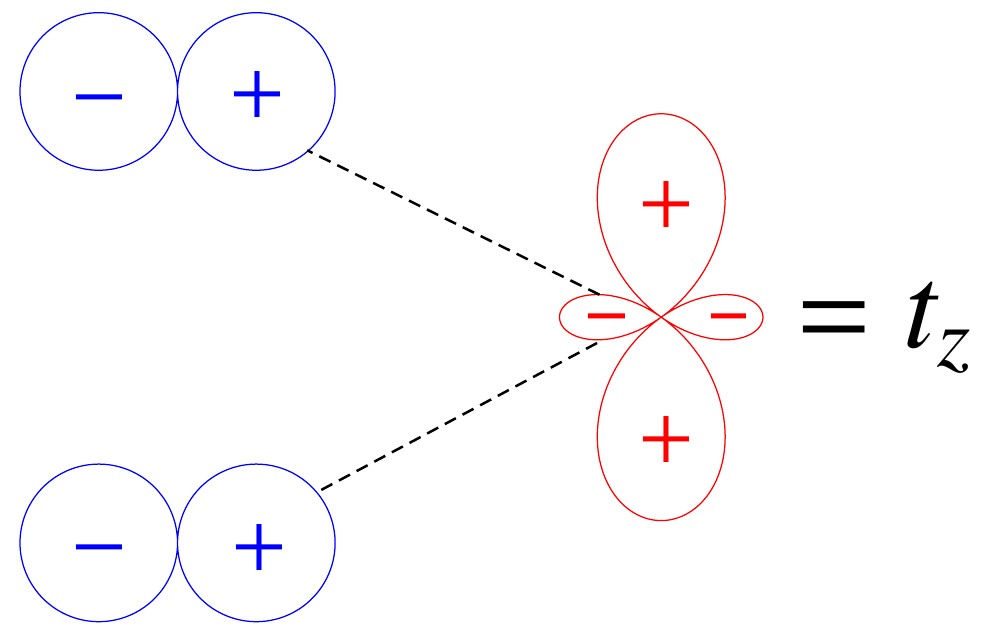}
\caption{Hopping integrals considered in the tight-binding model.}
\label{fig:hoppings}
\end{figure}
\begin{center}
\begin{table}
\begin{tabular}{|c|c|}
\hline
Tight-binding parameter&Value (eV)\\
\hline
\hline
$\Delta_0$&-1.512\\
\hline
$\Delta_2$&-3.025\\
\hline
$\Delta_p$&-1.276\\
\hline
$V_{pd\sigma}$&-2.619\\
\hline
$V_{pd\pi}$&-1.396\\
\hline
\end{tabular}
\caption{Tight-binding parameters considered in the calculation. The values for MoS$_2$ are taken from Ref.~\onlinecite{tb}.}
\label{tab:tb_parameters}
\end{table}
\end{center}
\begin{figure}
\includegraphics[width=0.6\textwidth]{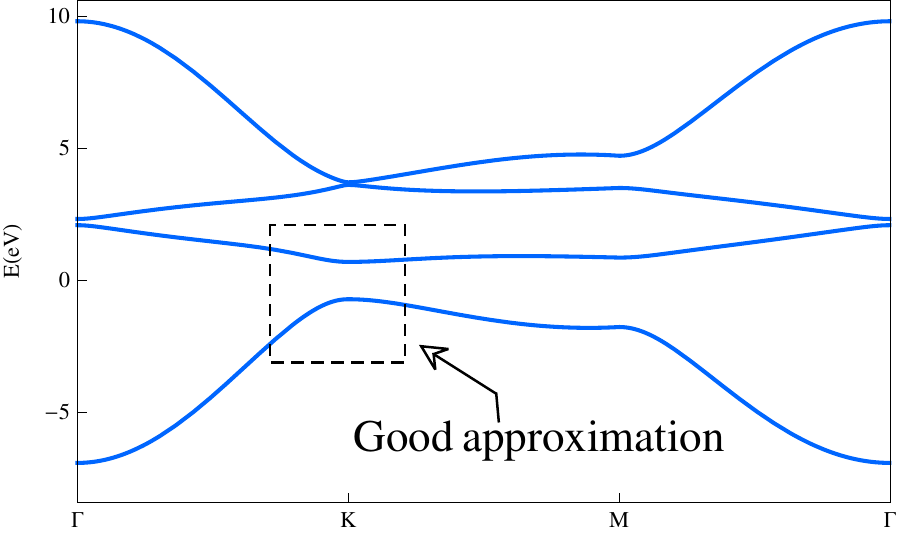}
\caption{Bands calculated within the tight-binding model described in the text. The model is only valid around $\mathbf{K}_{\tau}$ points (highlighted in the figure). We take the values summarized in Table~\ref{tab:tb_parameters} for MoS$_2$.}
\label{fig:bands}
\end{figure}
This model for the case of MoS$_2$ leads to the electronic bands shown in Fig.~\ref{fig:bands}.

\subsubsection{$\mathbf{k}\cdot\mathbf{p}$ theory}

At $\mathbf{K}_{\tau}$ points the Hamiltonian in first quantization reads as the matrix:\begin{equation}
\left(\begin{array}{cccc}
\Delta_0&0&3t_z&0\\
0&\Delta_2&0&-3\left(t_x+t_y\right)\\
3t_z&0&\Delta_p&0\\
0&-3\left(t_x+t_y\right)&0&\Delta_p
\end{array}\right)
\end{equation}
By diagonalizing this Hamiltonian we obtain the eigenvectors of the Bloch states maximally localized at $d_{3z^2-r^2}$ and $d_{x^2-y^2}+i\tau d_{xy}$ respectively, which define the conduction and band states respectively:\begin{eqnarray}
\left|\psi_c\right\rangle=\frac{1}{\sqrt{1+\left|c\right|^2}}\left(\begin{array}{c}
1\\
0\\
c\\
0
\end{array}\right)\nonumber\\
\left|\psi_v\right\rangle=\frac{1}{\sqrt{1+\left|v\right|^2}}\left(\begin{array}{c}
0\\
1\\
0\\
v
\end{array}\right)
\end{eqnarray}
where:\begin{gather}
c=-\frac{6t_z}{\sqrt{36t_z^2+(\Delta_p-\Delta_0)^2}+\Delta_p-\Delta_0}
\nonumber\\
v=\frac{6\left(t_x+t_y\right)}{\sqrt{36\left(t_x+t_y\right)^2+(\Delta_p-\Delta_2)^2}+\Delta_p-\Delta_2}
\end{gather}
Then, the $\mathbf{k}\cdot\mathbf{p}$ Hamiltonian is just:\begin{equation}
\mathcal{H}_{\mathbf{k}\cdot\mathbf{p}}=\left(\begin{array}{cc}
\left\langle\psi_c\right|\mathcal{H}_{TB}\left|\psi_c\right\rangle&\left\langle\psi_c\right|\mathcal{H}_{TB}\left|\psi_v\right\rangle\\
\left\langle\psi_v\right|\mathcal{H}_{TB}\left|\psi_c\right\rangle&\left\langle\psi_v\right|\mathcal{H}_{TB}\left|\psi_v\right\rangle
\end{array}\right)
\end{equation}
If we write $\vec{q}=\mathbf{K}_{\tau}+\vec{k}$ and expand in powers of $\vec{k}$ up to first order we get the Hamiltonian of Eq.~(1) of the main text with:\begin{gather}
\Delta=\frac{1}{2}\left(\Delta_0-\Delta_2-\sqrt{36 tz^2 + \left(\Delta_0 - \Delta_p\right))^2}-\sqrt{36 \left(t_x+t_y\right)^2+ \left(\Delta_2 - \Delta_p\right))^2} \right)=1.41\mbox{ eV}\nonumber\\
t=\frac{\sqrt{3}\left[c\left(tx-t_y\right)+vt_z\right]}{2\sqrt{1+\left|c\right|^2}\sqrt{1+\left|v\right|^2}}=1.19\mbox{ eV}
\end{gather}

\subsubsection{Coupling with strain}

We repeat this calculation by considering the change in the hopping integrals $t_{x,y,z}$ due to the displacement of the atoms. For the hopping integral between atoms at sites $\alpha$, $\beta$ we have:
\begin{equation}
t_{\alpha\beta}\rightarrow t_{\alpha\beta}+\frac{\partial t_{\alpha\beta}}{\partial\mathbf{r}}\left(\mathbf{u}_\alpha-\mathbf{u}_\beta\right)
\end{equation}
where $\mathbf{u}_{\alpha(\beta)}$ is the displacement of the atom at site $\alpha$ ($\beta$). For small displacements we can approximate:
\begin{eqnarray}
\frac{\partial t_{\alpha\beta}}{\partial\mathbf{r}}\approx \frac{\sqrt{3}}{a}t_{\alpha\beta}\beta_{\alpha\beta}\hat{\delta}_{\alpha\beta}\nonumber\\
\mathbf{u}_\alpha-\mathbf{u}_\beta\approx \frac{a}{\sqrt{3}}\hat{\delta}_{\alpha\beta}\cdot\vec{\partial}\mathbf{u}
\end{eqnarray}
where $\beta_{\alpha\beta}\equiv-\partial\ln t_{\alpha\beta}/\partial\ln a$ is the Gr\"uneisen parameter associated to $t_{\alpha\beta}$. Therefore, in the the previous matrix elements we have to consider now:
\begin{equation}
t_{x,y,z}\rightarrow t_{x,y,z}\left(1+\beta_{x,y,z}\delta_i\delta_ju_{ij}\right)
\end{equation} where $u_{ij}$ are the compnents of the strain tensor. By repeating the previous calculation at $\mathbf{K}_{\tau}$ we obtain the Hamiltonian of Eq.~(2) of the main text with:
\begin{align}
t\beta_0 &=\frac{3}{2}\left(\frac{c\beta_zt_z}{1+\left|c\right|^2}-\frac{v\left(\beta_xt_x+\beta_yt_y\right)}{1+\left|v\right|^2}\right)\nonumber\\
t\beta_1 &=\frac{3}{2}\left(\frac{c\beta_zt_z}{1+\left|c\right|^2}+\frac{v\left(\beta_xt_x+\beta_yt_y\right)}{1+\left|v\right|^2}\right)\nonumber\\
t\beta_2&=\frac{3\left(c\beta_xt_x-c\beta_yt_y-v\beta_zt_z\right)}{4\sqrt{1+\left|c\right|^2}\sqrt{1+\left|v\right|^2}}
\end{align}

\subsection{Gaps induced by superlattice potentials and Haldane-Kane-Mele phase}

We next provide the details of how a superlattice can be used to induce the subbands conduction and valence bands of a doped TMDCs. We approximate the bands of the homogeneous system by the two-band continuum-limit Hamiltonian introduced above, whose eigenvalues and eigenfunctions read:
\begin{gather}
 \epsilon_{\vec{\bf k}} = \pm \sqrt{\Delta^2 + ( v| \vec{\bf k} | )^2} \nonumber \\
 \left| \vec{\bf k} \right> = \left( \begin{array}{c}\cos \left( \frac{\theta_{\mathbf{k}}}{2} \right) \\ \sin \left( \frac{\theta_{\mathbf{k}}}{2} \right) e^{i \phi_{\mathbf{k}}} \end{array} \right) 
\end{gather}
where $\theta_{\mathbf{k}} = \arctan [ ( v_F | \vec{\bf k} | ) /\Delta]$ and $\phi_{\mathbf{k}} = \arctan ( k_y / k_x )$.

A superlattice potential hybridizes states $\left| \vec{\bf k} \right>$ and $\left| \vec{\bf k} + \vec{\bf G} \right>$ where the vectors $\vec{\bf G}$ define the superlattice. We consider the six lowest vectors $\vec{\bf G}$, with $| \vec{\bf G} | = ( 4 \pi ) / ( \sqrt{3} L )$, where $L = N a$ is the lattice constant of the $N \times N$ superlattice. We assume that  $v | \vec{\bf G} | \ll \Delta$ and that the superlattice potential, $V_{\vec{\bf G}}$, is such that $V_{\vec{\bf G}} \ll ( v | \vec{\bf G} | )^2 / ( 2 \Delta )$, so that perturbation theory in $V_{\vec{\bf G}}$ applies. We also assume that the lattice potential is sufficiently smooth, $| \vec{\bf G} | \ll | \vec{\bf K}_+ - \vec{\bf K}_- |$, where $\vec{\bf K}_{\pm}$ are the corners of the Brillouin Zone of the TMDC lattice, and neglect intervalley scattering.

Using first order perturbation theory, each set of three points at corners of the Brillouin Zone connected by superlattice reciprocal vectors leads to a $3 \times 3$ matrix:
\begin{align}
{\cal H}_{\kappa,\kappa'} &\equiv \left( \begin{array}{ccc} \epsilon_{\kappa,\kappa'} \pm \bar{v} k_x &V_{\kappa,\kappa'} &V_{\kappa,\kappa'}^*  \\ V_{\kappa,\kappa'}^* &\epsilon_{\kappa,\kappa'} \pm \bar{v} \left( - \frac{k_x}{2} +\frac{\sqrt{3} k_y}{2} \right) &V_{\kappa,\kappa'} \\ V_{\kappa,\kappa'} & V_{\kappa,\kappa'}^* & \epsilon_{\kappa,\kappa'} \pm \bar{v} \left( - \frac{k_x}{2} - \frac{\sqrt{3} k_y}{2} \right) \end{array} \right)
\label{hamil}
\end{align}
where $\epsilon_{\kappa,\kappa'} = \epsilon_0 = ( v | \vec{\bf K} | )^2 / ( 2 \Delta)$, $\bar{v} \approx ( v^2 | \vec{\bf K} | ) / \Delta$, and the two signs correspond to the $\kappa$ and $\kappa'$ points. For $V_{\kappa,\kappa'} = \left|V_{\kappa,\kappa'}\right| e^{i\phi_{\kappa,\kappa'}}$, the  the energies and eigenfunctions at the $\kappa$ and $\kappa'$ points, in the basis used to write eq.(\ref{hamil}) are:
\begin{align}
\epsilon_a &= \epsilon_0 + 2 \left|V_{\kappa,\kappa'}\right|\cos\left(\phi_{\kappa,\kappa'} \right) & \left|  a \right\rangle = \frac{1}{\sqrt{3}} \left( \left| 1 \right\rangle + \left| 2 \right\rangle + \left| 3 \right\rangle \right) \nonumber \\
\epsilon_b &= \epsilon_0 + 2 \left|V_{\kappa,\kappa'}\right|\cos\left(\frac{2\pi}{3}+\phi_{\kappa,\kappa'} \right) & \left|  b \right\rangle = \frac{1}{\sqrt{3}} \left( \left| 1 \right\rangle + e^{2 \pi i / 3} \left| 2 \right\rangle + e^{- 2 \pi i / 3} \left| 3 \right\rangle \right) \nonumber  \\
\epsilon_c &= \epsilon_0 + 2 \left|V_{\kappa,\kappa'}\right|\cos\left(\frac{4\pi}{3}+\phi_{\kappa,\kappa'}\right)  & \left|  c \right\rangle = \frac{1}{\sqrt{3}} \left( \left| 1 \right\rangle + e^{- 2 \pi i / 3} \left| 2 \right\rangle + e^{2 \pi i / 3} \left| 3 \right\rangle \right)
\label{solution}
\end{align}
In the case of $\phi_{\kappa,\kappa'}=0,\pi$, then $V_{\kappa,\kappa'}$ real, an expansion in powers of $| \vec{\bf k} |$ shows that states $| b \rangle$ and $| c \rangle$ define an effective $2 \times 2$ Dirac Hamiltonian with velocity $\bar{v} / 2$. For $\phi_{\kappa,\kappa'}=0$ this Dirac point gives the lowest edge of the highest valence subband, and for $\phi_{\kappa,\kappa'}=\pi$ the upper edge of the lowest conduction band. The degeneracy of these Dirac points is lifted for complex values of $V_{\kappa,\kappa'}$. The problem is equivalent to a gapped Dirac equation with gap $\Delta_{\kappa,\kappa'} =2\sqrt{3} | V_{\kappa,\kappa'} | \sin\left(\phi_{\kappa,\kappa'}\right)$. If $\phi_{\kappa'}$ and $\phi_{\kappa'}$ have different signs, the two gaps also have  opposite signs, leading to a lowest subband with a Chern number equal to one. This is a realization of Haldane's model~\cite{Haldane}.

\begin{figure}
\includegraphics[width=0.5\columnwidth]{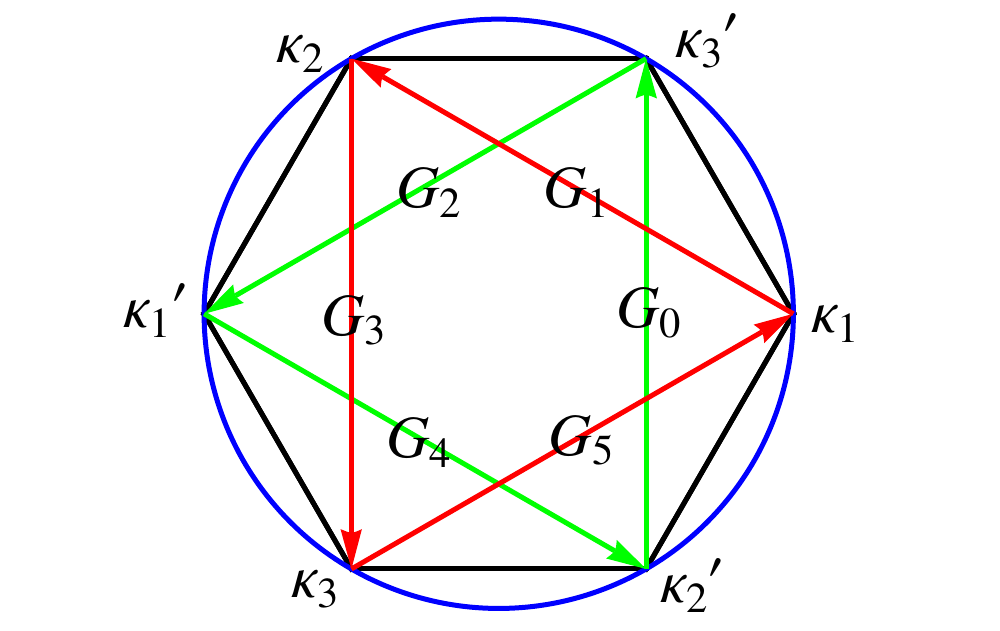}
\caption[fig]{\label{sketch} (Color online). Sketch of the effect of the Brillouin Zone in a superlattice, and states mixed by the superlattice potential.}
\end{figure}

The superlattice potential is a $2 \times 2$ matrix in the space span by conduction and valence band states, and can be divided into scalar, mass and vector components, which, in turn, can be even or odd under spatial inversion. Following Ref.~\onlinecite{falko} we define the functions:
\begin{gather}
f_1\left(\mathbf{r}\right)=\sum_{m=0...5}e^{i\mathbf{G}_m\cdot\mathbf{r}}\nonumber\\
f_2\left(\mathbf{r}\right)=i\sum_{m=0...5}(-1)^me^{i\mathbf{G}_m\cdot\mathbf{r}}\nonumber\\
\end{gather}
Then, we can construct the inversion-symmetric superlattice potentials as:
\begin{gather}
V_{s}= v\left|\mathbf{G}\right|\Delta_{s}f_1\left(\mathbf{r}\right)\nonumber\\
V_{m}=v\left|\mathbf{G}\right|\Delta_{m}f_1\left(\mathbf{r}\right)\sigma_z\nonumber\\
V_{g}=v\Delta_{g}\left(\sigma_x,\tau_z\sigma_y\right)\cdot\left(\hat{z}\times\nabla\right)f_2\left(\mathbf{r}\right)
\end{gather}
The coefficients $\Delta_{s,m,g}$ in these expressions 
are dimensionless phenomenological constants with the energy scale set by $v\left|\mathbf{G}\right|=\frac{4\pi t}{\sqrt{3}N}$.

The scalar potential has matrix elements:\begin{align}
\left< \mathbf{k}+\mathbf{G}_m \right| V_s ( \mathbf{G}_m ) \left| \bf{k} \right> &= v\left|\mathbf{G}\right|\Delta_s \left[ \cos \left( \frac{\theta_{\mathbf{k}+\mathbf{G}_m}}{2} \right) \cos \left( \frac{\theta_{\mathbf{k}}}{2} \right)+ \sin \left( \frac{\theta_{\mathbf{k}+\mathbf{G}_m}}{2} \right) \sin \left( \frac{\theta_{\mathbf{k}}}{2} \right)   
e^{i \left( \phi_{\mathbf{k}} - \phi_{\mathbf{k} +\mathbf{G}_m} \right)}
\right]
\end{align}
Equivalently for the mass potential:\begin{align}
\left< \mathbf{k}+\mathbf{G}_m \right| V_m ( \mathbf{G}_m ) \left| \bf{k} \right> &= v\left|\mathbf{G}\right|\Delta_m \left[ \cos \left( \frac{\theta_{\mathbf{k}+\mathbf{G}_m}}{2} \right) \cos \left( \frac{\theta_{\mathbf{k}}}{2} \right)- \sin \left( \frac{\theta_{\mathbf{k}+\mathbf{G}_m}}{2} \right) \sin \left( \frac{\theta_{\mathbf{k}}}{2} \right)   
e^{i \left( \phi_{\mathbf{k}} - \phi_{\mathbf{k} +\mathbf{G}_m} \right)}
\right]
\end{align}
The edges of the first subband are determined by the shifts in the energies of the corners of the superlattice Brillouin Zone. We assume that $|\mathbf{k} | = | \mathbf{k} + \mathbf{G}_m | = \kappa= ( 4 \pi ) /( 3 N a )$. Then:
\begin{align}
\left< \mathbf{k}+\mathbf{G}_{m} \right| V_{s,m} ( \mathbf{G}_m ) \left| \bf{k} \right> &\approx
 v\left|\mathbf{G}\right|\Delta_{s,m}\left[1-\frac{v^2\kappa^2}{4\Delta^2}\left(1\mp e^{i \left( \phi_{\mathbf{k}} - \phi_{\mathbf{k} +\mathbf{G}_m} \right)}\right)\right]
\end{align}
For the gauge potential we have:
\begin{gather}
\left< \mathbf{k}+\mathbf{G}_m \right| V_g ( \mathbf{G}_m ) \left| \bf{k} \right> =i\left(-1\right) ^{m} v\left|\mathbf{G}\right|\Delta_g \left[ \cos \left( \frac{\theta_{\mathbf{k}+\mathbf{G}_m}}{2} \right) \sin \left( \frac{\theta_{\mathbf{k}}}{2} \right)e^{i \left( \phi_{\mathbf{k}} - \phi_{\mathbf{G}_m} \right)}- \sin \left( \frac{\theta_{\mathbf{k}+\mathbf{G}_m}}{2} \right) \cos \left( \frac{\theta_{\mathbf{k}}}{2} \right)   
e^{i \left( \phi_{\mathbf{G}_m} - \phi_{\mathbf{k} +\mathbf{G}_m} \right)}
\right]\nonumber\\
\approx \left(-1\right) ^{m}\frac{v\kappa}{\Delta}v\left|\mathbf{G}\right|\Delta_g 
e^{i \frac{\phi_{\mathbf{k}} - \phi_{\mathbf{k} +\mathbf{G}_m}}{2}}\cos\left(\frac{m\pi}{3}-\frac{\phi_{\mathbf{k}} + \phi_{\mathbf{k} +\mathbf{G}_m}}{2}\right)
\end{gather}

The same can be done with the inversion-asymmetric superlattice potentials, defined as:\begin{gather}
\tilde{V}_{s}= v\left|\mathbf{G}\right|\tilde{\Delta}_{s}f_2\left(\mathbf{r}\right)\nonumber\\
\tilde{V}_{m}=v\left|\mathbf{G}\right|\tilde{\Delta}_{m}f_2\left(\mathbf{r}\right)\sigma_z\nonumber\\
\tilde{V}_{g}=v\tilde{\Delta}_{g}\left(\sigma_x,\tau_z\sigma_y\right)\cdot\left(\hat{z}\times\nabla\right)f_1\left(\mathbf{r}\right)
\end{gather}
By repeating the same calculation we obtain:\begin{gather}
\left< \mathbf{k}+\mathbf{G}_{m} \right| \tilde{V}_{s,m} ( \mathbf{G}_m ) \left| \bf{k} \right> \approx
i\left(-1\right)^m v\left|\mathbf{G}\right|\tilde{\Delta}_{s,m}\left[1-\frac{v^2\kappa^2}{4\Delta^2}\left(1\mp e^{i \left( \phi_{\mathbf{k}} - \phi_{\mathbf{k} +\mathbf{G}_m} \right)}\right)\right]
\nonumber\\
\left< \mathbf{k}+\mathbf{G}_m \right| \tilde{V}_g ( \mathbf{G}_m ) \left| \bf{k} \right>\approx -i\frac{v\kappa}{\Delta}v\left|\mathbf{G}\right|\tilde{\Delta}_g 
e^{i \frac{\phi_{\mathbf{k}} - \phi_{\mathbf{k} +\mathbf{G}_m}}{2}}\cos\left(\frac{m\pi}{3}-\frac{\phi_{\mathbf{k}} + \phi_{\mathbf{k} +\mathbf{G}_m}}{2}\right)
\end{gather}

From this analysis it is clear that inversion-asymmetric potentials are needed in order to induce a topological subband structure. To the leading order in $v\kappa/\Delta$, considering scalar potentials only, we have:
\begin{gather}
V_{\kappa}=\left\langle\kappa_1=\kappa_2+\mathbf{G}_4\right|V_s\left(\mathbf{G}_4\right)+\tilde{V}_s\left(\mathbf{G}_4\right)\left|\kappa_2\right\rangle=v\left|\mathbf{G}\right|\left(\Delta_s
+i\tilde{\Delta}_s\right)\nonumber\\
V_{\kappa'}=\left\langle\kappa_1'=\kappa_2'+\mathbf{G}_1\right|V_s\left(\mathbf{G}_1\right)+\tilde{V}_s\left(\mathbf{G}_1\right)\left|\kappa_2'\right\rangle=v\left|\mathbf{G}\right|\left(\Delta_s
-i\tilde{\Delta}_s\right)
\end{gather}
So $V_{\kappa,\kappa'}=v\left|\mathbf{G}\right|\sqrt{\Delta_s^2+\tilde{\Delta}_s^2}\times e^{\pm i\arctan\left(\frac{\tilde{\Delta}_s}{\Delta_s}\right)}$. Hence, the gap is $\Delta_{\kappa,\kappa'}=\pm2\sqrt{3}v\left|\mathbf{G}\right|\tilde{\Delta}_s$.
Either the lowest conduction or valence subbands derived from the bands at a given spin polarized valley in the band structure of the TMDC have a Chern number equal to one. This Chern number is compensated by the opposite value from the other valley due to time-reversal symmetry. Therefore, this system is effectively a realization of the Kane-Mele model~\cite{Kane_Mele}.

\end{widetext}


\begin{thebibliography}{30}
%
\bibitem{dical}
Q.~H. Wang, K. Kalantar-Zadeh, A. Kiss, J. N. Coleman,
and M. S. Strano, Nature Nanotech {\bf 7}, 699 (2012) and
references therein.
%
\bibitem{transistor}
B.~Radisavljevic, A.~Radenovic,~J. Brivio,
V. Giacometti, and A. Kis, Nature Nanotechnology,
{\bf 6}, 147 (2011);
H.~Nam, S.~Wi, H.~Rokni, M.~Chen, G.~Priessnitz, W. Lu,
and X. Liang, ACS Nano, {\bf 7}, 5870 (2013).
%
\bibitem{cite_qshe}
M.~Z. Hassan and C.~L. Kane, Rev.~Mod.~Phys. {\bf 82}, 3045 (2010);
X.-L. Qi and S.-C. Zhang, \emph{ibid} {\bf 83}, 1057 (2011); B. A. Bernevig
with T. L. Hughes, \emph{Topological Insulators and Topological Superconductors},
Princeton Univ. Press, (Princeton 2013),
and references therein.
%
\bibitem{qshe_app}
See e.g. talk by S.-C. Zhang, available from \url{http://www.aip.org/industry/ipf/2011/day2_8_Zhang/}
%
\bibitem{mercury_tell}
M. K\"onig, S. Wiedmann, C. Br\"une, A. Roth, H. Buhmann, L.~W.
Molenkamp, X.-L. Qi, S.C. Zhang, Science {\bf 308}, 766 (2007).
%
\bibitem{other}
I. Knez,  R.-R. Du, G. Sullivan, Phys. Rev. Lett. 107, 136603 (2011)
%
\bibitem{Weeks}
C. Weeks,  J. Hu, J. Alicea, M. Franz, and R. Wu,
Physical Review X {\bf 1}, 021001 (2011).
%
\bibitem{kane_mele}
C.~L. Kane and E.~J. Mele, Phys. Rev. {\bf 95}, 226801 (2013).
%
\bibitem{guinea_katsnelson_geim}
F. Guinea, M.~I. Katsnelson, and A.~K. Geim,
Nature Physics, {\bf 6}, 30 (2010).
%
\bibitem{crommie}
N. Levy \emph{et al.}
Science {\bf 329}, 544 (2010);
J. Lu,	 A.H. Castro Neto, and  K. P. Loh,
Nature Comm. {\bf 3}, 823 (2011).
%
\bibitem{vozmediano_katsnelson_guinea}
M.~A. Vozmediano, M. I. Katsnelson, and F. Guinea,
Physics Reports {\bf 496}, 109 (2010).
%
\bibitem{bernevig_zhang}
B.~A. Bernevig and S.-C. Zhang, Phys.~Rev.~Lett. {\bf 96}, 106802 (2006).
%
\bibitem{two_band}
D. Xiao, G.-B. Liu, W. Feng, X. Xu, and W. Yao,
Phys. Rev. Lett. {\bf 108}, 196802 (2012).
{\bf 88}, 045416 (2013).
%
\bibitem{tb}
E. Cappelluti, R. Rold\'an, J.A. Silva-Guillén,
P. Ordej\'on, F. Guinea, Phys. Rev. B {\bf 88}, 075409 (2013).
%
\bibitem{ochoa_roldan}
H. Ochoa and R. Rold\'an, Phys. Rev. B {\bf 87}, 245421 (2013).
%
\bibitem{so_dft}
Z. Y. Zhu, Y. C. Cheng, and U. Schwingenschl\"ogl, Phys. Rev. B {\bf 84}, 153402 (2011);
W. Feng, Y. Yao, W. Zhu, J. Zhou, W. Yao, and D. Xiao, \emph{ibid} {\bf 86}, 165108 (2012);
A. Korm\'anyos, V. Z\'olyomi, N.~D. Drummond,
P. Rakyta, G. Burkard, V.~I. Fal'ko,  \emph{ibid}
{\bf 88}, 045416 (2013).
%
\bibitem{epaps}
See supplementary material.
%
\bibitem{arpes}
W. Jin \emph{et al.} Phys. Rev. Lett. {\bf 111}, 106801 (2013);
S.~K. Mahatha, K.~D. Patel, abd K.~S.~R. Menon, J.~Phys. Cond. Matt.~
{\bf 24} 475504 (2012).
%
\bibitem{spinorbit}
K.~F. Mak \emph{et al.}
Phys. Rev. Lett.~{\bf 105}, 136805 (2010);
W. S. Yun \emph{et al.} Phys. Rev. B {\bf 85}, 033305 (2012).
%
\bibitem{beta}
We take $C>0$ and assume $\beta\approx 3$ according to M. Buscema \emph{et al.} ACS Nano Lett.{\bf 13}, 358 (2013).
%
\bibitem{dft}
R. C. Cooper, C. Lee, C.~A. Marianetti, X. Wei, J. Hone, and
J. W. Kysar, Phys. Rev. B {\bf 87}, 035423 (2013). The value of $\mu$ is obtained from the calculated Young's modulus $E$ and Poisson's ratio $\nu$ as $\mu=\frac{E}{2\left(1+\nu\right)}$.
%
\bibitem{eff_mass}
This is the effective mass inferred from the tight-binding model. The actual hole effective mass is slightly anisotropic~\cite{so_dft}. We take
$m^*= 0.5$ and neglect the anisotropy.
%
\bibitem{lacroix}
M. Taillefumier \emph{et al.} Phys. Rev. B {\bf 78}, 155330 (2008).
%
\bibitem{LGK11}
T. Low,  F. Guinea, and K. I.  Katsnelson, Phys. Rev. B {\bf 83} 195436 (2011).
%
\bibitem{S09}
I. Snyman, Phys. Rev. B {\bf 80}, 054303 (2009).
%
\bibitem{H88}
F. D. M. Haldane, Phys. Rev: Lett. {\bf 61}, 2015 (1988).
%
\bibitem{Yetal12}
M. Yankowitz {\it et al}, Nature Phys. {\bf 8}, 382 (2012).
%
\bibitem{Petal13}
L. A. Ponomarenko {\it et al}, Nature {\bf 497}, 594 (2013).
%
\bibitem{Detal13}
C. R. Dean {\it et al}, Nature {\bf 497}, 598 (2013).
%
\bibitem{Hetal13}
B. Hunt {\it et al}, Science {\bf 340}, 1427 (2013).
%
\bibitem{SN13}
O.~P. Sushkov, and A.~H. Castro Neto, Phys. Rev. Lett. {\bf 110}, 186601 (2013).
%
\bibitem{Aharonov}
Y. Aharonov and A. Casher,  Phys. Rev. A {\bf 19}, 2461 (1979);
M. Bander, Phys. Rev. B {\bf 41}, 9028 (1990).
%
\bibitem{note}
 $\Phi$ is assumed to be positive as in the uniform case. Note that $\chi(\mathbf{r})=-\int \frac{d\mathbf{r}^{\prime}}{2\pi}\ln(|\mathbf{r}-\mathbf{r}^{\prime}|)B_0(\mathbf{r}^{\prime})\sim- \frac{\Phi}{2\pi}\ln(|\mathbf{r}|)$ as $|\mathbf{r}|\rightarrow\infty$, and therefore $\psi\left(\mathbf{r}\right)\sim f\left(x+i\tau_z y \right)|\mathbf{r}|^{-\frac{\Phi}{\Phi_0}}$.
%
\bibitem{xu_moore}
C. Xu and J. Moore, Phys. Rev. B {\bf 73}  045322. (2006);
C. Wu, B. A. Bernevig, and S.-C. Zhang, Phys.~Rev.~Lett. {\bf 96},
106401 (2006).
%
\bibitem{roldan}
R. Rold\'an, E. Cappelutti, and F. Guinea, Phys. Rev. B {\bf 88},
054515 (2013).%
%
\bibitem{iridates}
A. Shitade \emph{et al.}, Phys. Rev. Lett. 102, 256403 (2009)
%
\end{thebibliography}

\begin{thebibliography}{5}

\bibitem{Slater-Koster}
J. C. Slater and G. F. Koster, Phys. Rev. {\bf 94}, 1498 (1954).

\bibitem{tb}
E. Cappelluti, R. Rold\'an, J.A. Silva-Guill\'en, 
P. Ordej\'on, F. Guinea, Phys. Rev. B {\bf 88}, 075409 (2013).

\bibitem{Haldane}
F. D. M. Haldane, Phys. Rev. Lett. \textbf{61}, 2015 (1988).

\bibitem{falko}
J. R. Wallbank, A. A. Patel, M. Mucha-Kruczyn'ski, A. K. Geim, and V. I. Fal'ko, Phys. Rev. B \textbf{87}, 245408 (2013).

\bibitem{Kane_Mele}
C. L. Kane and E. J. Mele, Phys. Rev. Lett. \textbf{95}, 226801, \textit{ibid}  146802 (2005).
%
\end{thebibliography}
\end{document}